\documentclass[prl,aps, twocolumn, nofootinbib,superscriptaddress]{revtex4-1}
\usepackage{amsmath}
\usepackage{amsfonts}
\usepackage{amssymb}
\usepackage{ragged2e}
\usepackage[font=small,labelfont=bf,justification=justified]{caption}
\usepackage{gensymb}
\usepackage{color}

  \makeatletter
    \renewcommand\@make@capt@title[2]{%
     \@ifx@empty\float@link{\@firstofone}{\expandafter\href\expandafter{\float@link}}%
      {\textbf{#1}}\@caption@fignum@sep#2\quad}%
    \makeatother
   
\makeatletter 
\renewcommand{\fnum@figure}{\textbf{Figure~\thefigure}}
\makeatother

\usepackage{amsmath}
\usepackage{amsfonts}
\usepackage{graphicx}
\usepackage{verbatim} 
\usepackage{subfig}

\begin{document}
\bibliographystyle{naturemag}

\title{Effective learning is accompanied by high dimensional and efficient representations of neural activity}

\author{Evelyn Tang}
\affiliation{Department of Bioengineering, University of Pennsylvania, PA 19104 USA}
\author{Marcelo G. Mattar}
\affiliation{Department of Psychology, University of Pennsylvania, Philadelphia, PA 19104 USA}
\author{Chad Giusti}
\affiliation{Department of Bioengineering, University of Pennsylvania, PA 19104 USA}
\author{Sharon L. Thompson-Schill}
\affiliation{Department of Psychology, University of Pennsylvania, Philadelphia, PA 19104 USA}
\author{Danielle S. Bassett}
\affiliation{Department of Bioengineering, University of Pennsylvania, PA 19104 USA}
\affiliation{Department of Electrical and Systems Engineering, University of Pennsylvania, PA 19104 USA}

\date{August, 2018}

\begin{abstract}	
A fundamental cognitive process is the ability to map value and identity onto objects as we learn about them. Exactly how such mental constructs emerge and what kind of space best embeds this mapping remains incompletely understood. Here we develop tools to quantify the space and organization of such a mapping, thereby providing a framework for studying the geometric representations of neural responses as reflected in functional MRI. Considering how human subjects learn the values of novel objects, we show that quick learners have a higher dimensional geometric representation than slow learners, and hence more easily distinguishable whole-brain responses to objects of different value. Furthermore, we find that quick learners display a more compact embedding of their neural responses and hence have a higher ratio of their task-based dimension to their embedding dimension --- consistent with a greater efficiency of cognitive coding. Lastly, we investigate the neurophysiological drivers of high dimensional patterns at both regional and voxel levels, and we complete our study with a complementary test of the distinguishability of associated whole-brain responses. Our results demonstrate a spatial organization of neural responses characteristic of learning, and offer a suite of geometric measures applicable to the study of efficient coding in higher-order cognitive processes more broadly.
\end{abstract}

\maketitle

\section{Introduction}

Essential to human cognition is the ability to group stimuli into meaningful identities. The emergence of such identities is accompanied by the development of a mapping encoded in the activity patterns of neural circuitry \cite{Freedman312}. Exactly how new information about objects is mapped into the correct groups, such that relevant information becomes associated, remains incompletely understood. Furthermore, it is not generally known how far apart such groups should be, and what kind of space efficiently embeds such a mapping. These concepts and questions are reminiscent of studies of coding efficiency in neural responses to low-level sensory stimuli \cite{articleBarlow,Atick:1990:TTE:1351041.1351047,OLSHAUSEN19973311,Lewicki2002} -- a notion quantifying a system's information processing given biophysical and metabolic constraints. An open question is whether similar principles of efficiency play a role in higher-level processes such as cognition \cite{Poldrack201512}. What goals and constraints must be balanced to enable such \textit{cognitive coding efficiency} \cite{Buzsaki2012,ZIMMER201244,HAIER1992415,Gold387,Heinzel1224}, and how might such efficiency support accurate perceptions and decisions? 

To formalize intuitive notions of space and organization in neural activity during the building of such mental maps, we use a geometric perspective adapted from machine learning \cite{Rigotti2013,DIEDRICHSEN2013225}. Specifically, we represent distributed neural responses as points in a multidimensional space. Applied to neuron-level data, such representations have been shown to be very effective in isolating an instrinsic low-dimensional subspace relevant to ongoing cognitive processes \cite{Rigotti2013,Ganguli200815,joshgold}. Here we extend these tools to the examination of large-scale neural responses in humans as they integrate information across many areas to form representations of novel objects \cite{doi:10.1093/cercor/1.1.1-a,doi:10.1146/annurev.neuro.27.070203.144220}, appreciate abstract properties of those objects \cite{Bartra2013412}, and both prepare and execute associated motor responses \cite{DIEDRICHSEN2013225}. Despite our growing understanding of the regions activated by such learning \cite{doi:10.1146/annurev.neuro.27.070203.144220,Bartra2013412,OpdeBeeck11796,Grill-Spector2014,COHEN200561}, a significant gap in knowledge lies in delineating how spatiotemporal patterns of neural responses in these activated regions allow for effective behavioral choices. Our approach complements multivoxel pattern analysis and related techniques -- which enable a local quantification of regional representations of objects or concepts \cite{kahnt2017decade,DIEDRICHSEN2013225} -- by offering tools that synthesize information across all brain regions simultaneously.

Fundamentally, these tools allow us to hypothesize that the dimension of a geometric representation of neural responses is related to the effective identification of stimuli and corresponding learned values. The simple intuition behind this hypothesis is that a higher dimension allows for an easier grouping of neural responses according to different objects in the geometric space. To test this hypothesis, we examine blood oxygen level dependent (BOLD) magnitudes at the regional and voxel levels, in a cohort of 20 healthy adult human subjects as they learn the values of twelve novel objects over the course of 4 consecutive days for a total of 80 experimental imaging sessions. Motivated by a desire to study parsimonious representations and also by recent work decoding object identity \cite{Waskom10743, Cichy2014,doi:10.1162/jocn}, stimulus response \cite{BZDOK2013381,Kahnt1493}, and markers of emotional and affective processing \cite{doi:10.1093/cercor/bhs065,doi:10.1093/scan/nsu089} from coarse-scale measurements across the brain, we spatially average these indirect measurements of neural activity in 83 regions of interest (ROIs) defined by a whole-brain anatomical parcellation. Next, we use a general linear model to deconvolve the hemodynamic response function to obtain approximate neural responses to each stimulus at the time point at which it was presented. We ask how the dimension of the geometric representation of such neural responses reflects the speed with which participants learn the objects' monetary values \cite{doi:10.1162/NETN_a_00021}. To answer this question, we study three aspects of the geometric organization of these neural responses: the task-based and embedding dimensions, and label assortativity.

We demonstrate that fast learners have higher dimensional \emph{task-based} geometric representations, allowing for an easier development of boundaries between neural responses to different stimuli. However, a potential disadvantage of using a high dimensional representation is that the brain might utilize more resources for the embedding of the information. To assess the presence or absence of this potential tradeoff, we study the \emph{embedding dimension}: the geometric representation of each subject's neural responses, with the map between stimulus and neural response shuffled uniformly at random. We find that the embedding dimension of a fast learner is more compact than that of a slow learner, suggesting that their neural responses form a more contained underlying subspace within the higher dimensional ROI space. The large ratio between the task-based dimension and the embedding dimension is indicative of efficient coding, and is observed most commonly in the participants who learned rapidly. To enhance our understanding of the anatomy driving these observations, we identify brain regions that most contribute to the emergence of high dimensional patterns in quick learners, and we further implement a voxel-level analysis to examine finer-scale structure in neural responses. Lastly, we use the complementary metric \emph{label assortativity} to characterize how easy it is to distinguish between neural responses. Our results confirm our prior analysis that fast learners have more distinguishable neural responses. Taken together, our approach provides novel insights into the geometry of neural responses supporting learning, and offers a suite of computational heuristics to intuitively describe cognitive processes more generally. 

\begin{figure}[tb]
\includegraphics[width=0.99\linewidth]{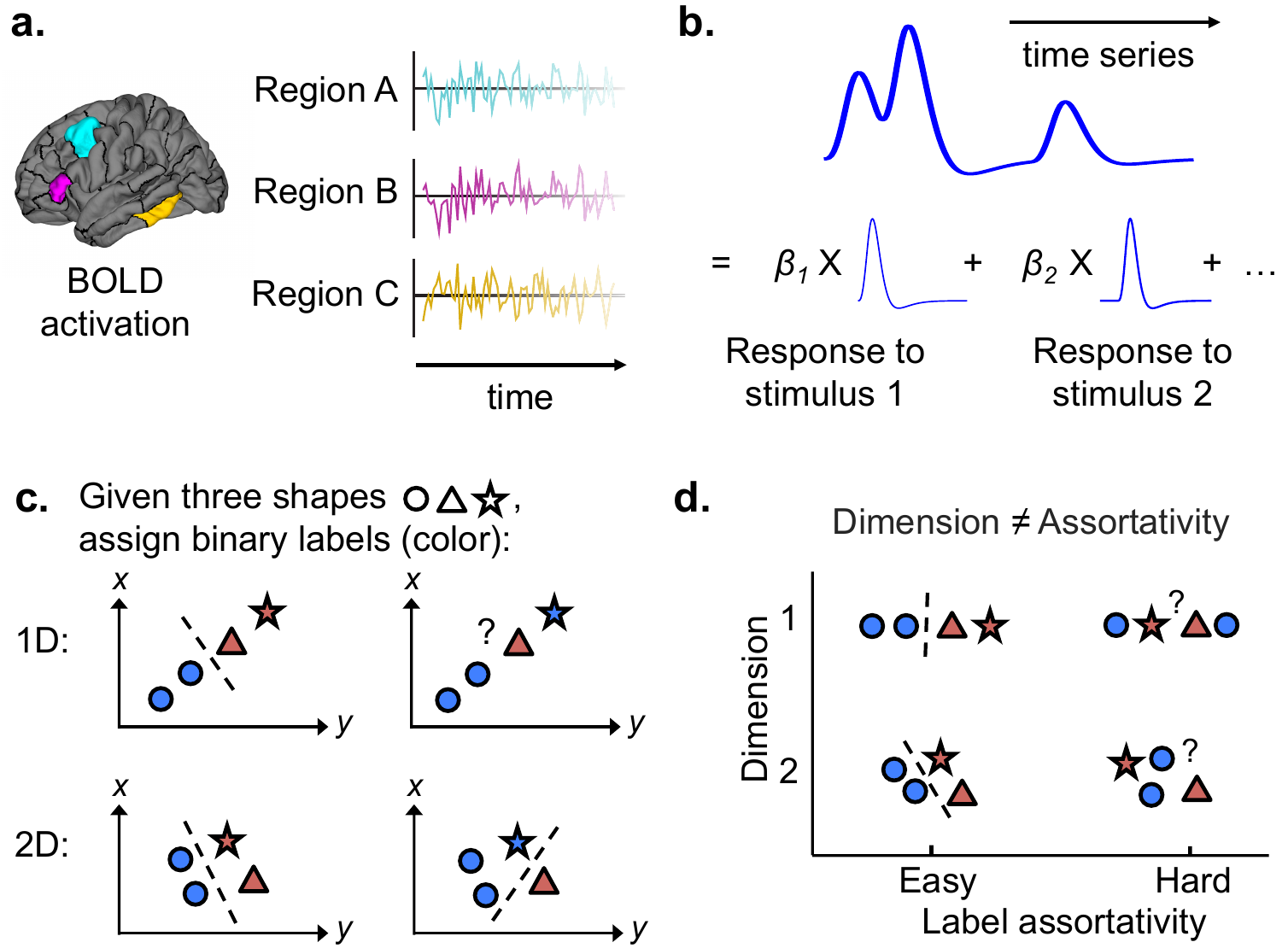}
\caption{\textbf{Neural responses from fMRI data; separability dimension and assortativity. a.} We measure the regional fMRI BOLD activation over 1 hour of task practice. \textbf{b.} Using a general linear model to deconvolve the hemodynamic response function from the BOLD time series, we obtain approximate neural responses, $\{\beta_i\}$, to each stimulus at the time point when it was presented. \textbf{c.} We assign binary labels (denoted by color) to the neural data (denoted by shapes). When the data are arranged in a low-dimensional manner (top row), some binary assignments will result in poorer separability, whereas in a higher dimension, these binary assignments can be more easily separated. The average performance of separability over different possible binary assignments gives the separability dimension; the axes $x$ and $y$ denote an ROI measurement space. \textbf{d.} Label assortativity does not depend strictly on dimension, and can measure a different geometric aspect of the same data. In panels \textbf{(c)} and \textbf{(d)}, dashed lines represent a classifier boundary, while question marks illustrate the difficulty of finding a clean boundary.} \label{fig:intro}
\end{figure}

\section{Results}

\subsection{Quick learners develop higher dimensional task-based representations of neural responses}

We seek to understand how the neural responses of subjects are distributed according to the task-relevant stimuli, and how this distribution reflects their learning ability. The dimensionality of the fMRI BOLD evoked responses (Fig. \ref{fig:intro}a-b) can be estimated based on the performance of a linear classifier in distinguishing assigned binary labels on the data \cite{Rigotti2013}. Intuitively, a given spatial arrangement of these responses will make it easier for any of the stimuli to be distinguished from the others, when the data are arranged in a higher-dimensional manner. Specifically, for $n$ stimuli there are $2^n$ ways to assign binary labels to these stimuli. When the data are arranged in a low-dimensional manner, some binary assignments will result in poorer separability, whereas in a higher dimension, these binary assignments will result in higher separability on average (see Fig. \ref{fig:intro}c). By exhaustively examining all $2^n-2$ choices of binary labelings and recording the resulting separability, the average performance over this combinatorial number of assignments yields the \textit{task-based} separability dimension (see Methods). An advantage of this process of averaging over many separating hyperplanes is a robustness of the results to noise: while the result in any particular hyperplane might be sensitive to perturbation, the average result will be stable. 

\begin{figure}[tb]
\includegraphics[width=0.99\linewidth]{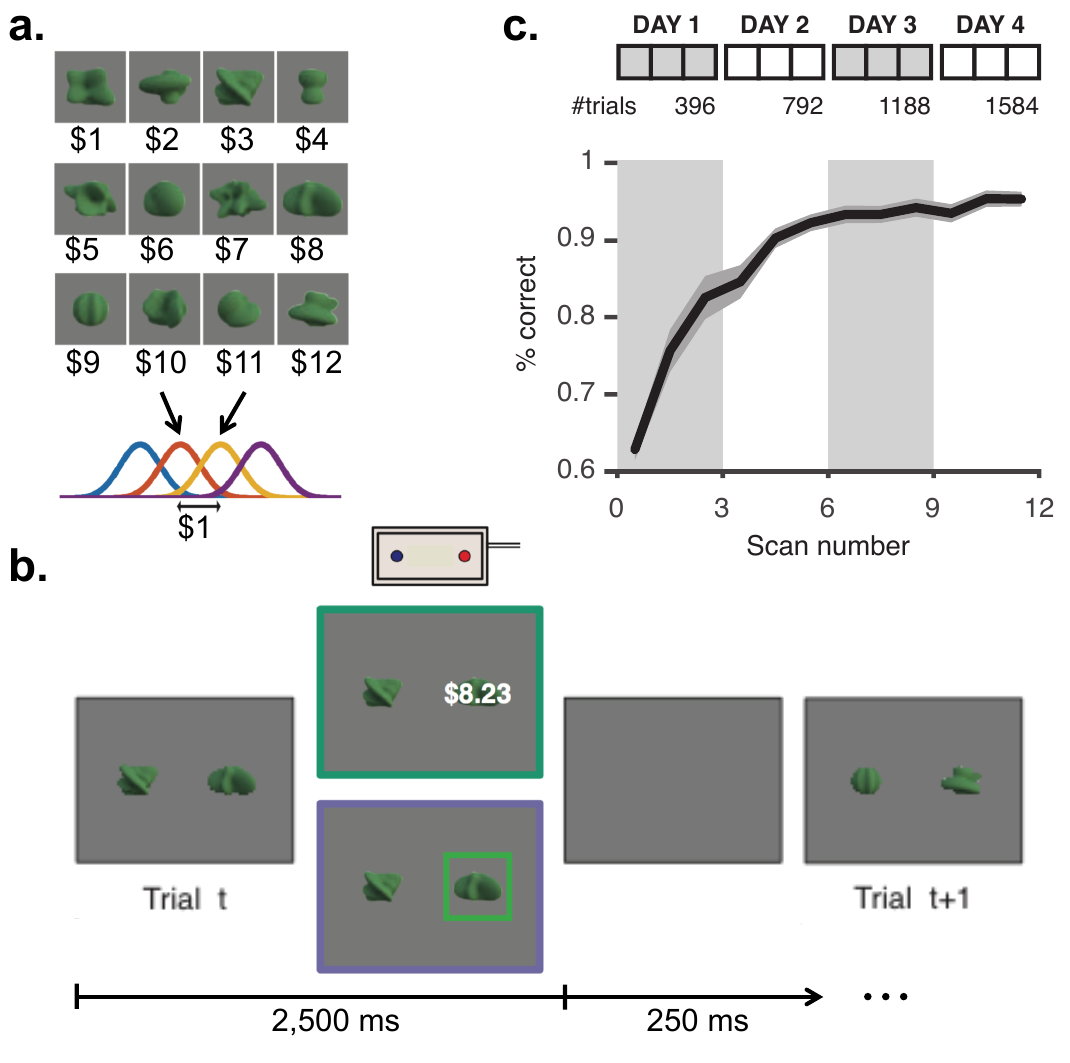}
\caption{\textbf{Experimental protocol and behavioral results. a.} Stimulus set and corresponding values. Twelve abstract shapes were computer-generated, and an integer value between \$1 and \$12 was assigned to each. On each trial, the empirical value of each shape was drawn from a Gaussian distribution with fixed mean, and standard deviation of \$0.50. \textbf{b.} Learning phase. Participants were presented with two shapes side-by-side on the screen and asked to choose the shape with the higher monetary value. Once a selection was made, feedback on their selection was provided. Each trial lasted 2.75 s (250 ms inter-stimulus interval). \textbf{c.} The experiment was conducted over four consecutive days, with learning phases (three experimental scans or 396 trials each day, for a total of 1584 trials) and a daily value judgement task where stimuli were presented singly. Participants' accuracy in selecting the shape with higher expected value improved steadily over the course of the experiment, increasing from chance level in the first few trials to approximately 95\% in the final few trials.}\label{fig:expt}
\end{figure}

We apply this method to the evoked neural responses of participants learning the value of twelve arbitrary computer-generated shapes (see Fig. \ref{fig:expt}a; \cite{doi:10.1162/NETN_a_00021}). Each shape was assigned a distinct and fixed monetary value. During the learning phase, participants were shown a pair of shapes simultaneously and asked to select which shape had the higher value, after which they received feedback based on their response (see Fig. \ref{fig:expt}b). This portion of the experiment was followed by a value judgement task, where participants were shown individual shapes and asked to indicate if the shape was one of the six least or one of the six most valuable shapes (see Methods). The learning phases and the value judgement task were repeated daily for four days (see Fig. \ref{fig:expt}c). As the sessions progressed, subjects improved in their abilities to select the shape with the higher expected value. By the conclusion of the second day of practice, all subjects reached a generally high level of performance (see Fig. \ref{fig:expt}c). To best distinguish individual differences in performance, we sought to identify the response accuracy from the value judgement session with the greatest individual variability, which provided a statistically rigorous metric and guided our subsequent analysis \cite{Cunningham2014}. We observed such greatest individual variability at the end of the first day, and we refer to this metric as the learning speed of participants.

\begin{figure*}[tb]
	\includegraphics[width=0.80\linewidth]{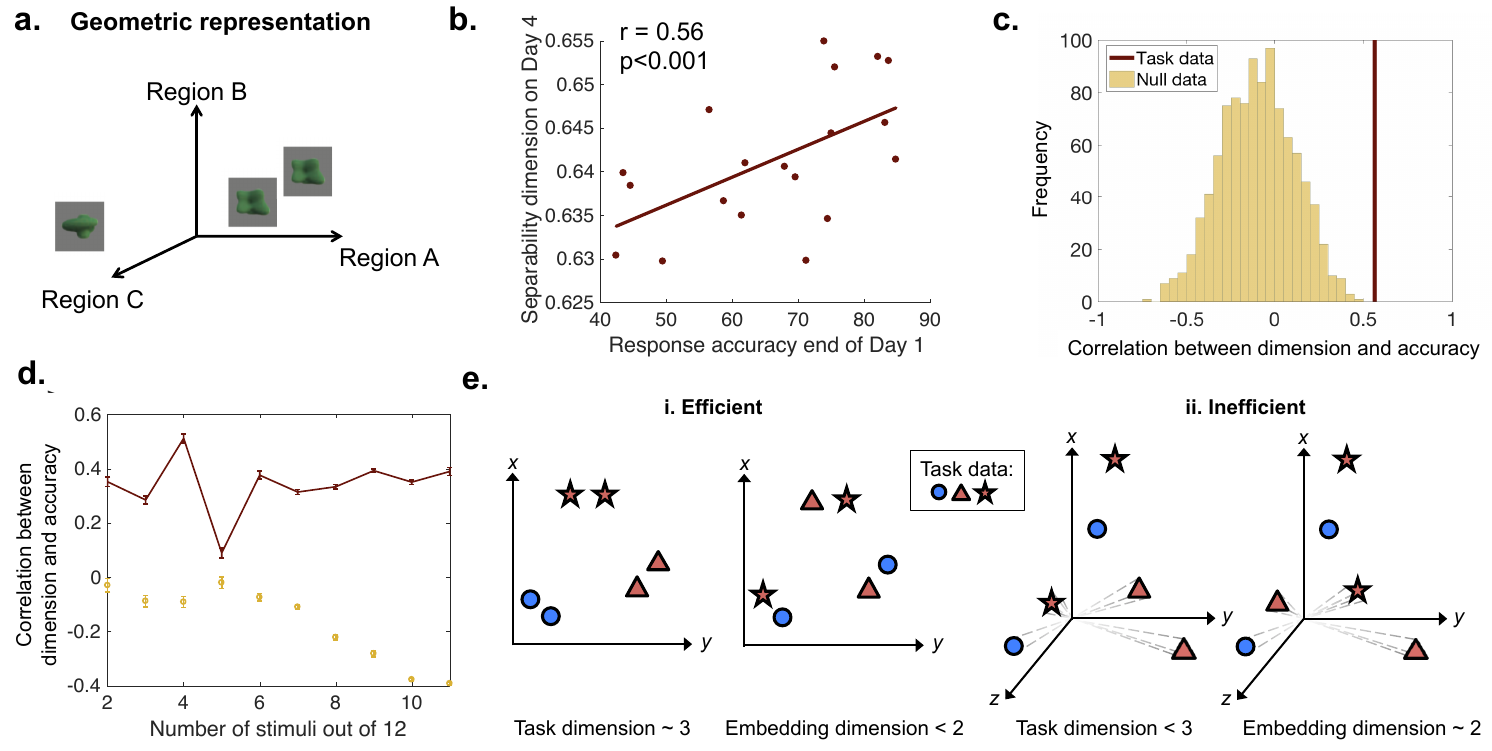}
	\caption{\textbf{Quick learners show higher dimensional and more efficient representations. a.} Schematic of how the presentation of each shape evokes neural responses across brain regions to contribute a data point in ROI space. Many shape presentations in a task session jointly form a point cloud or geometric representation, the dimension of which we can quantify. \textbf{b.} Relation between task-based separability dimension ($m=4$) and learning accuracy across 19 participants (Pearson's correlation coefficient $r=0.56$). We compare this correlation value with that observed in a permutation-based null model where object labels are shuffled uniformly at random, and the separability dimension is recalculated. We find that the true correlation is significantly greater than that expected under this null model with non-parametric $p<0.001$. \textbf{c.} Histogram of 1000 bootstrapped estimates of the value of the across-subjects correlation coefficient between subject-specific response accuracy and the dimension of subject-specific null data (gold bars). The correlation value estimated from the true task-based data is shown in red. \textbf{d.} Relation between the separability dimension and learning accuracy across subjects, for $m$ from 2 to 10. The true data is shown in red while the null data is shown in gold (using the same color scheme as in panel \textbf{c}). The estimates become more reliable as $m$ increases. We see that the true data displays a positive correlation with a magnitude far outside the error bars of the null model, which by constrast displays a negative correlation. These observations jointly suggest that fast learners have a large task-based dimension but small embedding dimension, overall forming an efficient representation of neural responses. \textbf{e.} Schematics of (i) an efficient representation with high task-based and low embedding dimensions, and of (ii) an inefficient representation with comparatively lower task-based and higher embedding dimensions. Our findings suggest that fast learners possess efficient neural representations, as manifest by a larger ratio of task-based to embedding dimensions; the axes $x$, $y$, and $z$ denote an ROI measurement space.}\label{fig:dim}
\end{figure*}

We seek to explore how the geometric representation of each subject's neural responses is related to their learning speed. To obtain this geometric representation, we use data from the value judgement task when shapes were presented one at a time and we apply a general linear model to obtain the neural response to each shape (see Fig. \ref{fig:intro}b), for every ROI. For each shape, the neural responses across all regions contribute a point in the ROI space. Hence, 140 shape presentations in one session jointly form a point cloud or geometric representation (see Fig. \ref{fig:dim}a), the dimension of which we quantify. Because we are most interested in understanding changes that occur directly due to learning over the full time course of the experiment, we investigate the task-based separability dimension of each subject that emerges by the end of the experiment: that is, on the fourth and final day of training. An alternative approach is to consider changes in the neural data from the first day to the fourth day. We also perform this analysis and report the results in the Supplement (see section ``Quick learners have an increasing dimension of representation across the experiment''; Fig. \ref{fig:neuralchanges}), which confirm our main findings. For $n=12$ however, calculating $2^n-2$ binary assignments is computationally expensive. Thus, in practice we choose a subset of $m=4$ stimuli over which to calculate this separability dimension. To ensure that our results do not depend on the particular subset of stimuli chosen, we repeat the calculation on 20 different combinations (roughly 7\%) out of the $\binom{n}{m}$ available choices, making sure that each shape was represented a roughly equal number of times throughout these sets. 

We find that the response accuracy of participants at the end of the first day of training is significantly correlated with their separability dimension at the end of the last day of training (Pearson's correlation coefficient $r=0.56$, see Fig. \ref{fig:dim}b). One subject was excluded due to errors during data collection (see Methods). To assess the statistical significance of this relation, we construct a null model by permuting the object labels of the neural responses uniformly at random, and then we calculate the separability dimension on these permuted data. Across subjects, we calculate the correlation between their response accuracy and the dimension of these null data in 1000 bootstrapped samples (gold bars in Fig. \ref{fig:dim}c). We observe that the true task-based data fall significantly outside this distribution with non-parametric $p<0.001$. This finding suggests that participants who learn more quickly display a larger task-based separability dimension of their representations, which allows for easier distinguishability between stimuli associated with different values. While we chose to use behavioral data with the largest individual variance for the purposes of statistical rigor, we note that this result also survives multiple hypothesis testing for behavioral accuracy data from all four days. In a sensitivity analysis, we also examine the separability dimension of neural responses from the other days, but we find that this correlation is the strongest with data from the final day (see Supplementary Information), suggesting that these higher dimensional responses emerge most clearly over time and the course of the experiment.

\subsection{Quick learners have a lower embedding dimension and hence overall more efficient representations}
 
Intuitively, a high-dimensional response provides flexibility in coding for task-based information but naturally uses more resources and has more potential to be distorted by errors. In contrast, a low-dimensional response uses less resources and has less potential for error than higher dimensional representations in the encoding process. How might quick learners potentially balance these two competing factors to develop efficient neural responses? To address this question, we extend our calculations from the previous section across a range of values of $m$, the cardinality of the subset of shapes from which the dimension is estimated. We calculate the correlation between separability dimension and response accuracy for the true task-based data and for the null data in 100 bootstrapped samples, up to $m=10$ (see Fig. \ref{fig:dim}d; Methods). Firstly, we notice that the true data are consistently positively correlated (red points) and fall far outside the error bars of the null data (gold points), confirming that across a range of $m$ the true data reflect quick learners having a higher task-based dimension of their representations. In fact, the results at large $m$ are particularly instructive as the combinatorics of $ 2^m-2$ averaged over for each calculation lead to a strong convergence of the results as reflected in very small error bars. Lastly, we note that while the positive correlation between task-based dimension and learning accuracy holds over a range of $m$ values, given that $m=4$ provides the strongest signal and is relatively computationally feasible to calculate in large quantities, further investigations into the task-based dimension are done using $m=4$.

Across subjects, we further observe that the correlation between separability dimension and learning accuracy is negative in the null data, particularly for large $m$ (see Fig. \ref{fig:dim}d). Intuitively, these data are the geometric distribution of neural responses without object labels, and thus reflect the embedding space of neural activity during the task. We therefore refer to the separability dimension of these null data as the \textit{embedding dimension}. Surprisingly, the negative correlation between subjects' learning accuracy and embedding dimension shows that fast learners have a lower embedding dimension, complementing their higher task-based dimension. This large ratio of task-based dimension to embedding dimension for fast learners suggests an efficient cognitive coding: the use of a smaller amount of embedding resources from which a more informative set of task-relevant features can be constructed.  We provide a low-dimensional schematic comparing such geometric arrangements in Fig. \ref{fig:dim}e. While the use of efficiency as a construct in cognitive science has been debated \cite{Poldrack201512}, here we provide a mathematical definition that contrasts the coding for meaningful content with the neural activity involved \textit{per se}, via the ratio between task-based and embedding dimensions. 

\subsection{Quick learners show high dimensional task-based representations within local brain regions}

To better understand the main effects reported in the previous sections, we first seek to determine which regions contribute most to the higher task-based dimension observed in quick learners. To address this question, we conduct a virtual lesioning analysis in which we remove brain regions one at a time, and then we recalculate the separability dimension of the modified representation. The regions whose absence causes the largest change in the observed correlation between separability dimension and response accuracy across subjects were found to be the left hippocampus and right temporal pole, respectively (magnitude of $z$-score $>2$ or $p<0.023$, uncorrected; see Supplement for details).

\begin{figure}[tb]
\includegraphics[width=0.92\linewidth]{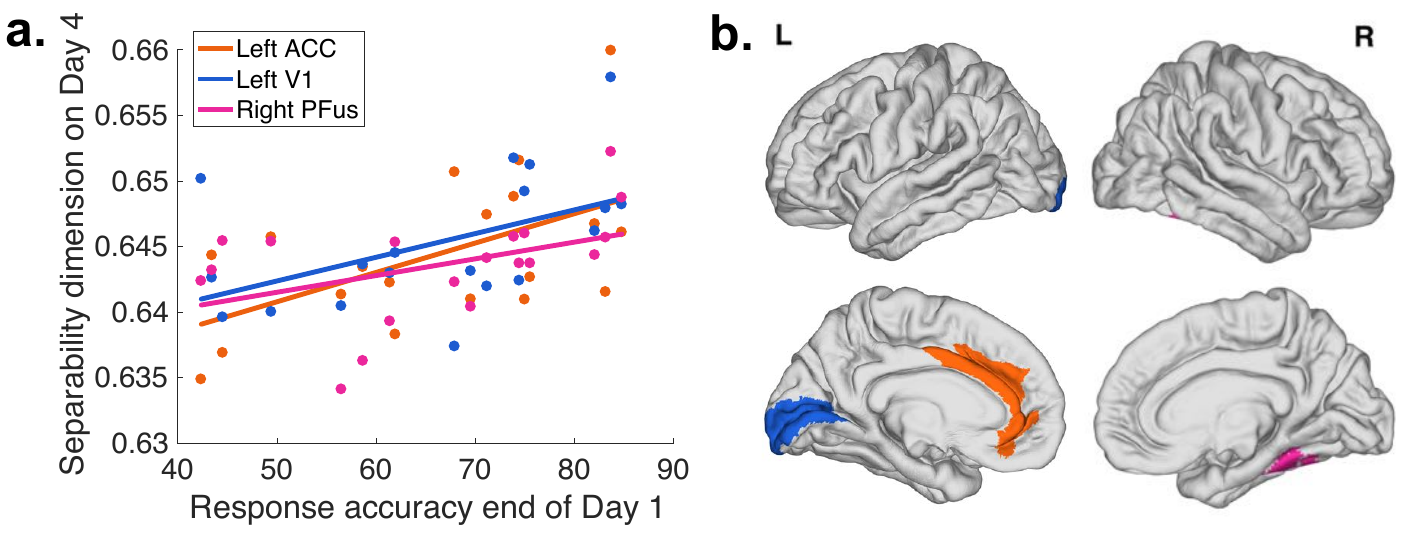}
\caption{\textbf{Quick learners show a larger dimension of responses in certain task-relevant regions at the voxel level. a.} We study regions of 300 (or fewer) voxels that we hypothesize to be involved in the processing of value and the learning of shapes. We see that three regions show a positive correlation between learning accuracy and separability dimension, with non-parametric $p\leq0.05$ compared to the null model of shuffled data. \textbf{b.} Topographical representation of these three regions on the surface of the brain: the left anterior cingulate cortex, left primary visual area, and right posterior fusiform. We note that the laterality of this latter effect is consistent with prior work demonstrating that the right and left posterior fusiform exhibit differential responses during object recognition \cite{Vuilleumier2002,KOUTSTAAL2001184,SIMONS2003613}. }\label{fig:voxel}
\end{figure}

In addition, we have up to this point studied neural activity across the whole-brain and the separability dimension of such neural activity. It is natural to ask if this relationship between learning ability and the dimension of neural responses can also be found in the multivoxel patterns of single brain regions hypothesized to be relevant for task performance. To address this question, we adapt our approach to examine ten regions of interest composed of 300 (or fewer) voxels (see Methods and Supplement). Following the prior analyses, we examine the correlation between separability dimension in the neural data in each local region and the participants' learning accuracy. Overall, we note that none of the regions show a negative correlation between their separability dimension and learning accuracy. Moreover, we find that three regions show a significant positive correlation, greater in magnitude than expected in the null model of shuffled data (non-parametric $p\leq0.05$; see Fig. \ref{fig:voxel}): the left anterior cingulate and primary visual cortices, as well as the right posterior fusiform cortex. We note that only the non-parametric test for the left anterior cingulate displayed $p\leq0.05$ after correcting for multiple comparisons (see Table \ref{tab:voxels}). Notably, the anterior cingulate cortex is thought to play a role in reward-based learning \cite{bush2002dorsal}, while the visual areas V1 and posterior fusiform are involved in the representation of lower-level and higher-level features of objects, respectively \cite{grill2003neural}.  Our findings therefore suggest that these regions are comparatively more engaged in the creation of a value-related heuristic at a local level. 

\begin{table}[!hb]
 \centering
\caption {\textbf{Brain regions where a higher dimensional representation is correlated with learning ability.} Pearson's correlation coefficient values and non-parametric $p$-values are given from comparison with the null model. The left anterior cingulate passes $p<0.005$ corrected for multiple comparisons (marked with $^*$).} 
\begin{tabular}{|c|c|c|c|c|}
\hline
No. of voxels & Brain region & Hemisphere & $r$ & $p$  \\
\hline
300& Anterior cingulate & Left &  0.54 & 0.003$ ^*$ \\
300 & Primary visual  & Left &  0.49  &   0.016 \\
300 & Posterior fusiform & Right & 0.61& 0.050 \\
\hline
\end{tabular} \label{tab:voxels}
\end{table}

\subsection{Quick learners develop more assortative representations}

\begin{figure}[tb]
\includegraphics[width=0.99\linewidth]{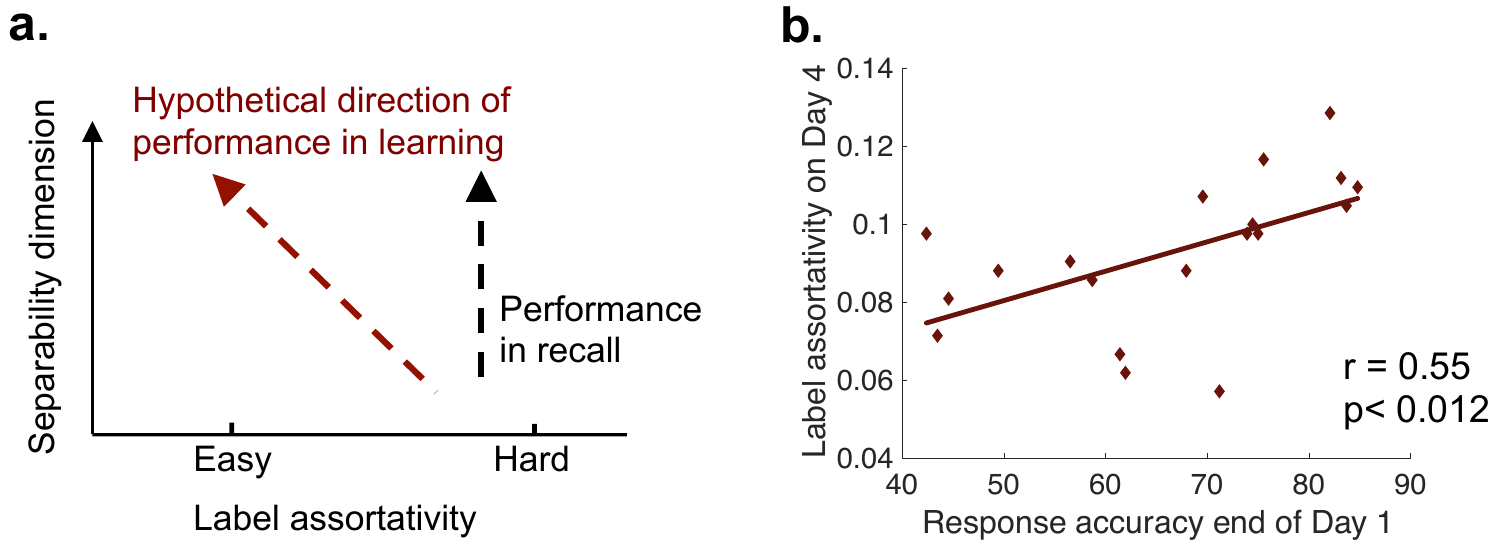}
\caption{\textbf{Dimension and assortativity provide a geometric depiction of neural data. a.} As different cognitive processes can exhibit typified geometric changes in the neural responses to various stimuli, we hypothesize that learning performance is associated with both higher dimension and higher assortativity. \textbf{b.} The distinct metric of label assortativity (according to all labels; see Fig. \ref{fig:intro}d) across the whole brain, shows that quick learners display a higher assortativity ($r=0.55$; red markers), compared to the shuffled data in gold with non-parametric $p=0.012$.} \label{fig:sep}
\end{figure}

Besides separability dimension, a complementary geometric measure is that of \textit{label assortativity}, which simply identifies how easily distinguishable the neural responses are from each other according to all labels, and not just binarized labels. While data that are more assortative are typically also higher dimensional, it is also possible for these metrics to vary independently (see Fig. \ref{fig:intro}d). Data can be arranged in a high or a low dimensional manner but still be easily classifiable (Fig. \ref{fig:intro}d, left) or data can be arranged in a high or low dimensional manner but be difficult to classify (Fig. \ref{fig:intro}d, right). Hence, the analysis of both metrics provides distinct and potentially independent information regarding the organization of the data. We hypothesize that quick learners should show a more assortative representation, in addition to having a higher task-based dimension (see Fig. \ref{fig:sep}a). Here, we calculate assortativity using a linear support vector machine, chosen because of its simple interpretability. When examining the same neural data from the value judgement session at the end of the fourth day, we find a positive correlation between their assortativity and the response accuracy of participants on the first day. Comparing this correlation to that observed in the null model in which labels are randomly permuted, we find that this correlation of $r=0.55$ is significant with non-parametric $p=0.012$ (see Fig. \ref{fig:sep}b). Intuitively, these data suggest that participants who learn more quickly have a more assortative pattern of neural responses than participants who learn less quickly. To verify that the metrics of separability dimension and label assortativity do not have a strict overlap, we note that one metric explains approximately $r^2=34\%$ of the variance of the other metric.

\section{Discussion}

Here we develop and apply a novel computational framework to reveal how the high-dimensional neural responses of quick learners allow for greater distinguishability of meaningful stimuli while requiring fewer informational resources. Our observations are enabled by emerging methods from machine learning and data science \cite{Rigotti2013,DIEDRICHSEN2013225}, which can be used to estimate the instrinsic dimension of a representation despite pervasive measurement noise. We extend the metric of the task-based dimension \cite{Rigotti2013} to study a complex cognitive task in whole-brain neural data, and we also introduce the new idea of the embedding dimension. In a cohort of 20 healthy adult humans learning the value of novel objects over the course of four days, we find that participants who learn most quickly display uniquely optimized neural responses to encode the cognitive processes associated with the task. The joint profile of the task-based and embedding dimensions allows us to quantify a concept of cognitive coding efficiency, based on the ratio between these two dimensions for each individual. We complement this examination with supporting studies of finer neuroanatomy (assessing multivoxel patterns) and computation (assessing local assortativity). Broadly, our work offers a suite of tools to characterize response geometry, thereby offering a simple and intuitive explanation for how individuals learn to successfully distinguish between relevant stimuli in their environment over time.
~\\
~\\
\noindent \emph{A notion of cognitive coding efficiency.} The concept of coding efficiency has been exercised at smaller spatial scales to characterize the (often unexpectedly low) dimension of neural representations. For example, neuronal spiking patterns measured in the lateral intraparietal area as macaques engage in a visual spatial attention task maps onto a one-dimensional dynamical trajectory \cite{Ganguli200815}. The simplicity and low-dimensionality of these dynamics marks disparate cognitive processes from decision-making and attentional shifting, to biased representations that arise from associative learning \cite{joshgold}. Indeed, such low-dimensionality is almost ubiquitous in neuronal measurements \cite{Cunningham2014,Machens350,SSolla}, although this often saturates the low dimensional bound set by the limited complexity of neural tasks commonly used today \cite{Gao214262}, or their autocorrelation structure \cite{FUSI201666}. Within this low-dimensional manifold, temporal variation in this ``effective'' dimension of neural activity can also indicate temporal variation in behavior \cite{Sadtler2014}. For example, as macaques engage in a recall task, the estimated task-based dimension from neural spiking activity in the prefrontal cortex is higher during correct responses than during incorrect responses \cite{Rigotti2013}.

Extending previous methods, we introduce two complementary types of dimension (task-based and embedding) that allow insight into learning capacity and cognitive flexibility. Our results are consistent with the notion that the substantially different use of these two types of dimension allows the efficient encoding of contextually relevant data, potentially supporting optimal learning strategies. The compression of a large amount of information or content into a restricted number of channels has been studied in other cognitive domains such as sensory processing \cite{articleBarlow,Atick:1990:TTE:1351041.1351047,OLSHAUSEN19973311,Lewicki2002}. In light of these historical contributions, our results suggest that similar principles of geometric efficiency may extend to higher-order cognitive processes in humans. Further work could directly investigate commonalities in such principles across different scales of space and time. Such an investigation is in principle made possible by the fact that while the absolute value of these geometric metrics depends on the particular measurement technique, relative changes in value could be used to compare between data collected \emph{across} wholly different measurement techniques. 
~\\
~\\
\noindent \emph{Complex cognitive tasks require new models of cognitive coding efficiency.} Recent theoretical studies use biologically plausible models to demonstrate that complex tasks such as image recognition or sensory processing, are supported by high dimensional representations, which in turn allow for an accurate readout of stimulus identity \cite{Barak3844,BABADI20141213,Litwin-Kumar2017}. These and other theoretical developments show that the two types of dimension (task-based and embedding dimension) may have very different advantages and behavior, even within the same experiment or within the same neural network \cite{PhysRevX.8.031003}. An efficient balance between these two types of dimension may control a generalization-discrimination trade-off \cite{Barak3844}, and new models accounting for these two dimensions are necessary especially for the fundamental understanding of complex cognitive tasks. In a separate line of work, the concept of efficiency has been applied to large-scale human neuroimaging data, predominantly to describe situations where the behavior of subjects appears similar but neural activation is greater for one group (which is taken to be the ``less efficient'' group) than for the other \cite{ZIMMER201244,HAIER1992415,Gold387,Heinzel1224}. For instance, in an experiment involving working memory, less neural activity was needed for trained items as compared to new items \cite{ZIMMER201244}. The authors interpret this difference as a correlate of a gain in neural efficiency, and that training causes a more efficient neural representation. However, it has been pointed out that this interpretation does not shed light on the relationship between these two facts \cite{Poldrack201512}. In our study, we show that a more compact dimension of neural activation is simultaneously tied to larger information content in the same neural activation, leading to the idea of efficiency in the representation itself. This notion is more akin to how the concept of efficiency is used in other contexts in the neuroscience literature, such as in studies of efficient coding in sensory systems \cite{articleBarlow,Atick:1990:TTE:1351041.1351047,OLSHAUSEN19973311,Lewicki2002} or in studies of network efficiency \cite{Bullmore2012,10.1371/journal.pcbi.1000748}, where a maximal amount of information is conveyed through a fixed (or smaller) feature or basis set. That the efficient cognitive coding we observe also appears differentially in subjects who learn faster is consistent and intuitive, but is not in itself required for our definition of efficiency. Hence our calculations of the dimension of representation provide a rigorous framework for quantifying and reasoning about the efficiency of cognitive coding, which can be measured and compared in other cognitive processes.

In our experiment, subjects were presented with a set of shapes designed to have no visual features that correlated with their monetary value (see Methods). Each subject was required to flexibly reassign new values to these shapes through the course of the experiment. In general, humans can be guided to act according to what has been previously reinforced, or to move towards promising sources of future reward \cite{10.2307/1884852, SHIZGAL1997198,doi:10.1146/annurev.neuro.29.051605.112903}. Our work examines the neural basis that supports this flexible identification of new value to existing objects, and how such objects become distinguished from each other in the representation of neural activity according to task-based cues. Indeed, upon investigating data from sessions where subjects are asked to evaluate the size and not the value of each shape, fast learners show no particular difference in the dimension of their neural responses (see Supplementary Information). Our results complement previous investigations into the relevance of cognitive flexibility for effective learning \cite{10.2307/2785779,Bassett03052011} and the underlying processes of executive function \cite{shine,Braun15092015}, while illuminating the emergent geometric architecture of the neural responses of effective learners. Future work could study if efficient geometric representations arise in individuals who exhibit higher degrees of cognitive flexibility and dynamic reorganization of neural responses.
~\\
~\\
\noindent \emph{Changes in neural representations during learning and practice.} In seeking to decipher the rules of adaptation, learning, and development, it is common to examine how neural mechanisms support or foster behavioral patterns. In a complementary perspective, one can examine how temporally localized decisions or short term behaviors can drive adaptation or change in neural circuitry and long term habits \cite{behaviorchange}. This latter perspective has motivated studies identifying changes in fMRI measurements of brain activity following video game playing associated with improvements in visuospatial and attention-related skills \cite{10.1371/journal.pone.0189110}, as well as in the rate of regional subcortical glucose metabolism \cite{HAIER1992134}. Changes from tasks involving spatial navigation and visuomotor coordination have also been identified in structural brain properties, with effects outlasting even a short intensive gaming period \cite{MOMI201862}. In our study, as subjects learn to associate rewards of different magnitudes to novel stimuli, it is likely that new neural representations would emerge to represent these distinct groups.
~\\
~\\
\noindent \emph{Role of single regions within a broader whole-brain geometry.} Geometry and topology can be investigated across multiple scales of any complex system or its emergent dynamics \cite{betzel2016multi}. While some systems can display heterogeneity in geometric principles across spatial and temporal scales, others display greater scale-invariance, with the principles at one scale being recapitulated at other scales \cite{Khaluf20170662}. Applying our methods at different scales, we find that neural activity patterns elicited by value judgments of learned stimuli display similar geometric principles whether assessed at the level of the whole brain, or at the level of multivoxel patterns in single brain areas. Our choice to begin with an analysis of ROIs across the whole brain complements prior studies that often focus on fine-grained voxel patterns, and captures global organization which would be relevant during value learning. On a smaller scale we find that the left primary visual and anterior cingulate cortices, and right posterior fusiform of quick learners display a differential increase in dimension. While we focus on just ten local regions, each of which are hypothesized to play an important role in the cognitive processes elicited by this task, it would be of interest to expand the study to additional regions or sets of regions defined with other methods. Then, using the computational techniques that we introduce, one could begin to bridge the regional drivers of whole-brain simplicity and complexity in response geometry.
~\\
~\\
\noindent \emph{Methodological considerations.} We note that there are several methodological considerations that are pertinent to our study. First, while the GLM extracts neural responses from the time series averaged across entire regions, it could also be useful to perform this extraction on time series at the voxel level before averaging, which may also decrease noise from irrelevant signals. Second, dimension and assortativity constitute starting points for a deeper analysis, and further work could identify the exact topology of the response. Third, the broad geometric methods that we develop and utilize here could be complemented by a dynamical study to assess how this geometry evolves across time. Fourth, while our cohort of twenty subjects already demonstrates significant evidence for geometric features that distinguish quick from slow learners, these results could well be verified across larger samples. Fifth, in our work, we find a significantly higher dimension in the neural responses of quick learners on the last day (more than in the previous days, see Supplementary Information), suggesting that this higher-dimensional and more efficient representation emerges most clearly over time and training. However, our results remain correlative and cannot suggest a causal link between this high-dimensional representation and effective learning. 

Finally, we study a single cognitive task, and future work could extend these notions to other cognitive domains during different experiments, or as different cognitive processes are engaged. In a previous experiment examining recall performance in trained macaques, the two estimates of dimension and decoding accuracy (analogous to assortativity) are differentially related to behavior \cite{Rigotti2013}. Specifically, while the dimension of the macaque's neural representation was predictive of the macaque's performance, the decoding accuracy of the same neural data instead remained constant in both error and correct trials. These observations raise fundamental questions about whether different cognitive processes can exhibit typified geometric changes in the neural responses. In humans, a particularly interesting context in which to study such differences is the mental states engendered by ``explore" \emph{versus} ``exploit" behaviors common in general human experience \cite{Addicott2017}, which are thought to give rise to diffuse \emph{versus} structured neural representations.
~\\
~\\
\noindent \emph{Conclusion.} Here we offer a computational framework for quantifying and understanding the geometry of neural responses in humans. The tools that we develop and exercise hold promise for the analysis of other complex cognitive tasks due to their general applicability to non-invasive neuroimaging and notable robustness to noise. We illustrate the utility of these tools in characterizing the organization of neural activity associated with effective cognitive performance and efficient cognitive coding during the learning of abstract values associated with novel objects. Our results suggest that effective learners are marked by a type of cognitive coding efficiency characterized by high-dimensional geometric representations in concert with a compact embedding of the task-based information. Our observations motivate future work in cognitive and clinical neuroscience examining the generalizability of this notion of efficiency, and its relevance for disease.
 
~\\
~\\
\paragraph{Acknowledgments}
We thank Brett Falk for helpful discussions and Sarah Solomon for helpful comments on an earlier version of this manuscript. This work was supported by the John D. and Catherine T. MacArthur Foundation, the Alfred P. Sloan Foundation, the Army Research Laboratory and the Army Research Office through contract numbers W911NF-10-2-0022 and W911NF-14-1-0679, the National Institute of Health (2-R01-DC-009209-11, 1R01HD086888-01, R01-MH107235, R01-MH107703, R01MH109520, 1R01NS099348 and R21-M MH-106799), the Office of Naval Research, and the National Science Foundation (BCS-1441502, CAREER PHY-1554488, BCS-1631550, and CNS-1626008). The content is solely the responsibility of the authors and does not necessarily represent the official views of any of the funding agencies.

\bibliography{dimension_bib,bibfile}

\newpage
~
\newpage
\section{Online Methods}
\subsection{Dimension estimation}
Given several types of data such as the shapes in Fig. \ref{fig:intro}c and given that there can be several measurements for the same shape, we can assign a binary label to each shape, here represented by the color. In our case, each shape represents a neural response to one of $n$ stimuli. Given $n$ stimuli or shapes, there are $2^n$ ways to assign binary labels to these data. We can then ask how separable are these binary groups, across all $2^n-2$ relabellings \cite{Rigotti2013}. Note the additional $-2$ is because 2 cases out of the $2^n$ assign the same label to all of the data, and it is clearly unnecessary to calculate separability in those cases. We can see that when the data are arranged in one dimension, it becomes hard to separate the binary groups in all but one of the binary assignments. When the data are in a higher dimension, it will be easier to separate these binary groups. Hence the average binary separability over different assignments estimates the separability dimension of this geometric representation of the data, i.e. a higher value indicates that the data effectively live in a larger dimensional space. 

All simulations were performed in MATLAB (MathWorks). To calculate linear separability on the binary categories, we used the default validation scheme within the  \textit{Classification Learner} application, retaining the default option of 5-fold cross-validation. This algorithm partitions the data into 5 disjoint sets or folds, chosen randomly but with roughly equal size. For each fold, the algorithm trains a linear SVM using the out-of-fold observations, and then assesses model performance using in-fold data. Next, the average test error is calculated over all folds, to yield the separability for each binary assignment, which is a number between 0 and 1. We repeat this process over all binary assignments to obtain the separability in each case, and take the average separability over all $2^n-2$ assignments for $n$ types of data (or object identities). The resulting average is the \textit{separability dimension}, which has a monotonic relation with the cardinal dimension. Separability dimension is hence a useful proxy for cardinal dimension, and is sufficient to show relative differences between individuals, which is the purpose of our study. Note that the cardinal dimension (which is more intuitively familiar and $\geq1$) could be inferred by counting $N_c$, the number of successful binary assignments above a threshold, and relating that to the cardinal dimension $d$ using $d=\log_2 N_c$ \cite{Rigotti2013}. However, the number of data points needed to extrapolate this cardinal dimension ($\sim4000$ in \cite{Rigotti2013}) exceeds the amount of measured data available from typical experiments, and this number increases with task complexity. Hence, estimating the cardinal dimension often requires the introduction of additional data resampling techniques, which we chose not to use.

We perform this analysis on $m$ subsets of the stimuli. That is, for $m$ stimuli out of the 12 there are $\binom {12} {m}$ ways to assign binary labels. We choose 20 draws out of the different possible combinations in a uniform way, such that each stimuli is represented a similar number of times. This can be done for $m=2,...,10$, where $\binom {12} {m}>20$, and for $m=11$ we use all 12 possible draws. In order to preserve statistical rigor we do not study $m=12$ as there would be only one draw for $m=12$. For most calculations, we choose to use $m=4$ as a mid-size subset due to computational tractability, except in Fig. \ref{fig:dim}c where we show results for all $m<12$ to verify that the conclusions remain similar. 

\subsection{Linear SVM and cross validation}

In calculating binary separability, the MATLAB linear support vector machine (SVM) is used with cross-validation by partitioning the data in five folds. For each fold, a model was trained using the out-of-fold observations, after which model performance was assessed using in-fold data. The average test error is calculated over all folds to provide an estimate of the predictive accuracy of the final model, and is used as the measure of binary separability.  A similar cross-validation procedure is used to calculate label assortativity, where in this case the MATLAB linear SVM is also used with the data retaining all $n=12$ distinct labels.

\subsection{Value-learning experiment}

\subsubsection{Participants}
Twenty human participants (nine female; ages 19--53 years; mean age = 26.7 years) with normal or corrected vision and no history of neurological disease or psychiatric disorders were recruited for this experiment. All participants volunteered and provided informed consent in writing in accordance with the guidelines of the Institutional Review Board of the University of Pennsylvania (IRB \#801929). Participants had no prior experience with the stimuli or with the behavioral paradigm. 

\subsubsection{Stimuli design}

The novel stimuli were 3-dimensional shapes generated with a custom built MATLAB toolbox (code available at \href{http://github.com/saarela/ShapeToolbox}{http://github.com/saarela/ShapeToolbox}) and rendered with RADIANCE \citep{Ward:1994:RLS:192161.192286}. ShapeToolbox allows the generation of three-dimensional radial frequency patterns by modulating basis shapes, such as spheres, with an arbitrary combination of sinusoidal modulations in different frequencies, phases, amplitudes, and orientations. A large number of shapes were generated by selecting combinations of parameters at random. From this set, we selected twelve that were considered to be sufficiently distinct from one another. A different monetary value, varying from \$1.00 to \$12.00 in integer steps, was assigned to each shape (Fig. \ref{fig:expt}a). These values were uncorrelated with any parameter of the sinusoidal modulations, so that visual features were not informative of value.

\subsection{Experimental paradigm}

Subjects learned the monetary value of 12 novel visual stimuli over the course of four consecutive days \cite{doi:10.1162/NETN_a_00021}. Each day comprised of the following phases: (i) a size judgment task; (ii) a learning phase; (iii) a repetition of the size judgment task; (iv) a value judgment task. A 10-minute resting-state session preceded the experiments on each day. In the main text, we report data only from the value judgment task.

\subsubsection{Learning phase}

On each trial of the experiment, participants were presented with two shapes side-by-side on the screen and asked to choose the shape with the higher monetary value in an effort to maximize the total amount of money in their bank. Feedback (explicit or implicit) was given based on their response (Fig. \ref{fig:expt}b). The shape values on a given trial were independently drawn from a Gaussian distribution with mean equal to the true monetary value and the standard deviation equal to \$0.50 (Fig. \ref{fig:expt}a). This variation in the trial-specific value of a shape was incorporated in order to ensure that participants thought about the shapes as having worth, as opposed to simply associating a number or label with each shape. The average accuracy in selecting the shape with the highest mean value at each trial gradually improved over the course of the experiment, increasing from approximately 50\% (chance) in the first few trials to approximately 95\% in the final few trials.

\subsubsection{Value judgement task}
The value judgment task scans consisted of consecutive presentations of shapes drawn from the set (1500 ms presentation and 250 ms inter-stimulus interval) as participants indicated whether the shape was one of the six least or one of the six most valuable shapes. No feedback was given in this task.

We analyze data from the value judgement scans (both the BOLD data and participants' response accuracy) in the main text of the paper. We focus specifically on these data because the presentation of single stimuli in these sessions allows for the isolation of neural responses to each shape, which would be harder to disentangle from the simultaneous presentation of two shapes characteristic of the task used in the learning sessions. Note that the fMRI time series were poorly recorded for one participant in the value judgement session of the first day, due to a lack of synchronization between the computer and the scanner. Hence this participant was excluded from the analyses, with the other 19 subjects contributing data for the main analyses described in this paper. 

The behavioral data reported in the main text is the accuracy in this task (specifically, the accuracy at the end of the first day), while the neural data reported in the main text is measured from this task (specifically, based on day 4).

\subsubsection{Size judgement task}
The size judgment task scans consisted of consecutive presentations of shapes drawn from the set and presented with a $\pm$ 10\% size modulation (1500 ms presentation and 250 ms inter-stimulus interval) as participants indicated whether the shape was presented in a slightly larger or smaller variation.

\subsection{Image Acquisition}

We collected blood oxygen level dependent (BOLD) functional MRI data from each participant as they performed the task. 

\subsubsection{Learning phase}

A total of 12 scan runs over 4 days were completed by each person (three scans per session), totaling 1584 trials (Fig. \ref{fig:expt}c). Participants completed 20 min of the main task protocol on each scan session, learning the values of the 12 shapes through feedback. The sessions were comprised of three scans of 6.6 min each, starting with 16.5 seconds of a blank gray screen, followed by 132 experimental trials (2.75 sec each), and ending with another period of 16.5 seconds of a blank gray screen. Stimuli were back-projected onto a screen viewed by the participant through a mirror mounted on the head coil and subtended 4 degrees of visual angle, with 10 degrees separating the center of the two shapes. Each presentation lasted 2.5 sec (250 ms inter-stimulus interval) and, at any point within a trial, participants entered their responses on a 4-button response pad indicating their shape selection with a leftmost or rightmost button press. Stimuli were presented in a pseudorandom sequence with every pair of shapes presented once per scan.

\subsubsection{Value and size judgement tasks}
A total of 4 scan runs over 4 days were completed by each person (one scan per session) for the value judgement task, while a total of 8 scan runs over 4 days were completed by each person (two scans per session) for the size judgement tasks. Each scan lasted 5 minutes and 22 seconds (184 trials). Stimuli were back-projected onto a screen viewed by the participant through a mirror mounted on the head coil and sub-tended 4 degrees of visual angle. Each presentation lasted 1.75 seconds (250 ms inter-stimulus interval) and, at any point within a trial, participants entered their responses on a 4-button response pad indicating their shape selection with a leftmost (least valuable) or rightmost (most valuable) button press, during the value judgement tasks. During the size judgement tasks, these leftmost and rightmost button presses  corresponded to smaller and larger shapes respectively. Stimuli were presented in a counterbalanced sequence.

\subsection{MRI data collection and preprocessing}

Magnetic resonance images were obtained at the Hospital of the University of Pennsylvania using a 3.0 T Siemens Trio MRI scanner equipped with a 32-channel head coil. T1-weighted structural images of the whole brain were acquired on the first scan session using a three-dimensional magnetization-prepared rapid acquisition gradient echo pulse sequence (repetition time (TR) 1620 ms; echo time (TE) 3.09 ms; inversion time 950 ms; voxel size 1 mm $\times$ 1 mm $\times$ 1 mm; matrix size 190 $\times$ 263 $\times$ 165). A field map was also acquired at each scan session (TR 1200 ms; TE1 4.06 ms; TE2 6.52 ms; flip angle 60\degree; voxel size 3.4 mm $\times$ 3.4 mm $\times$ 4.0 mm; field of view 220 mm; matrix size 64 $\times$ 64 $\times$ 52) to correct geometric distortion caused by magnetic field inhomogeneity. In all experimental runs with a behavioral task, T2*-weighted images sensitive to blood oxygenation level-dependent contrasts were acquired using a slice accelerated multiband echo planar pulse sequence (TR 2,000 ms; TE 25 ms; flip angle 60\degree; voxel size 1.5 mm $\times$ 1.5 mm $\times$ 1.5 mm; field of view 192 mm; matrix size 128 $\times$ 128 $\times$ 80). In all resting state runs, T2*-weighted images sensitive to blood oxygenation level-dependent contrasts were acquired using a slice accelerated multiband echo planar pulse sequence (TR 500 ms; TE 30 ms; flip angle 30\degree; voxel size 3.0 mm $\times$ 3.0 mm $\times$ 3.0 mm; field of view 192 mm; matrix size 64 $\times$ 64 $\times$ 48). 

Cortical reconstruction and volumetric segmentation of the structural data was performed with the Freesurfer image analysis suite \citep{dale1999cortical}. Boundary-Based Registration between the structural image and the mean functional image was performed with Freesurfer \emph{bbregister} \citep{greve2009accurate}. Preprocessing of the resting state fMRI data was carried out using FEAT (FMRI Expert Analysis Tool) Version 6.00, part of FSL (FMRIB's Software Library, \href{http://www.fmrib.ox.ac.uk/fsl}{www.fmrib.ox.ac.uk/fsl}). The following pre-statistics processing was applied: EPI distortion correction using FUGUE \citep{jenkinson2004improving}; motion correction using MCFLIRT \citep{jenkinson2002improved}; slice-timing correction using Fourier-space time series phase-shifting; non-brain removal using BET \citep{smith2002fast}; grand-mean intensity normalization of the entire 4D dataset by a single multiplicative factor; highpass temporal filtering (Gaussian-weighted least-squares straight line fitting, with sigma=50.0s).

Nuisance time series were voxelwise regressed from the preprocessed data. Nuisance regressors included (i) three translation ($X, Y, Z$) and three rotation ($pitch, yaw, roll$) time series derived by retrospective head motion correction ($R=[X,Y,Z,pitch,yaw,roll]$), together with expansion terms ([$R$,$R^2$,$R_{t-1}$,$R_{t-1}^2$]), for a total of 24 motion regressors \citep{friston1996movement}); (ii) the first five principal components of non-neural sources of noise, estimated by averaging signals within white matter and cerebrospinal fluid masks, obtained with Freesurfer segmentation tools and removed using the anatomical CompCor method (aCompCor) \citep{behzadi2007component}; and (iii) an estimate of a local source of noise, estimated by averaging signals derived from the white matter region located within a 15 mm radius from each voxel, using the ANATICOR method \citep{jo2010mapping}. Global signal was not regressed out of voxel time series \citep{murphy2009impact, saad2012trouble, chai2012anticorrelations}. Instead, we follow recent guidelines by removing local white-matter signal and other non-neural sources  \citep{power2015recent,murphy2016towards}.

\subsection{GLM to extract stimuli responses from BOLD time series}

From the BOLD time series of 0.5 Hz, we interpolate the data to obtain a time series corresponding to the frequency of presentation of stimuli during the value judgment session (at 1.75 s intervals). We then use a general linear model (GLM) to obtain the static responses to each of these stimuli, $\{\beta_i\}$, for 184 stimuli in each sequence, see Fig. \ref{fig:intro}b. This procedure is repeated from each ROI, such that each stimulus has a $\beta_i$ from each of the 83 ROIs. Hence, each stimulus can be embedded as a point in the 83-dimensional ROI space. From here we keep the results for the first 140 stimuli shown in each session out of all 184 stimuli, which jointly form a 140-point data cloud or geometric representation in this ROI space, see Fig. \ref{fig:dim}a. This choice to truncate the data past 140 trials was dictated by the fact that the MRI acquisition does not continue past the length of the hemodynamic response for several of the last stimuli, thus providing inadequate data for GLM decoding.

\subsection{Whole-brain parcellation}

For the whole-brain analyses, we subdivide participants' gray matter volume into 83 cortical and subcortical areas in both hemispheres, based on regions assigned from the Lausanne atlas  \cite{10.1371/journal.pone.0048121}. For a replication of our results on a different whole-brain parcellation, please see the Supplement. 

\subsection{Voxel level study of brain regions}

We examine ten brain regions: posterior fusiform, anterior cingulate, orbitofrontal, lateral occipital, and primary visual cortices, each from the left and right hemisphere. We use the Group-Constrained Subject-Specific (GSS) method for defining the regions \cite{julian2012algorithmic}. For each region, a large parcel is defined based on an existing parcellation \cite{glasser2016multi}, within which a maximum of 300 voxels with highest object-\emph{versus}-scrambled $t$-statistic contrast from an independent localizer were selected. For lateral occipital and posterior fusiform, the parcels were downloaded from \href{http://web.mit.edu/bcs/nklab/GSS.shtml}{http://web.mit.edu/bcs/nklab/GSS.shtml}). This procedure allowed the selection of ROIs that exhibited univariate responses to objects in a subject-specific manner.

\subsection{Data availability}

The datasets generated during and analysed during the current study are available from the corresponding author on reasonable request.

~
\newpage

\section{Supplement}
\subsection{Regional drivers of the relationship between representation and behavior}

To better understand the main effects reported in the main text, we perform several \emph{post-hoc} analyses. Specifically, we seek to determine which regions contribute most to the higher task-based dimension observed in quick learners. To address this question, we conduct an exploratory analysis using a virtual lesioning approach in which we remove brain regions one at a time, and then recalculate the separability dimension of the modified representation. Here we report the regions whose absence causes the largest change in the observed correlation between separability dimension and response accuracy across subjects (magnitude of $z$-score $>2$ or $p<0.023$, uncorrected). Note that as this is a ranking procedure of which regions contribute the most, we simply report the regions with the largest deviation from the distribution of contributions from each region \cite{Honey10240}. This analysis does not lend itself to a correction for multiple comparisons, and is commonly used in examining which brain regions most strongly drive a particular effect \cite{Honey10240,VANDENHEUVEL2013683,10.1371/journal.pbio.0060159}.
\begin{figure}[h]
	\includegraphics[width=0.95\linewidth]{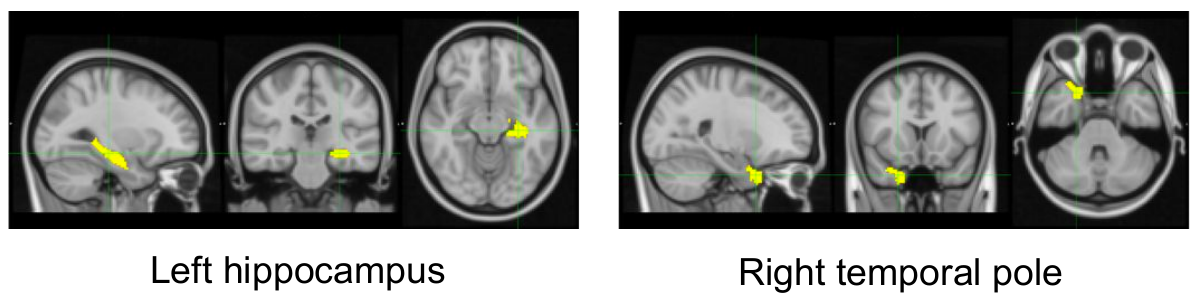}
	\caption{\textbf{Regional drivers of the relationship between representation and behavior.}  A virtual lesioning experiment shows the brain regions that most weaken the correlation between task-relevant separability dimension and learning accuracy upon removal ($z$-score $<-2$).}\label{fig:virtual}
\end{figure}

We find that removal of the left hippocampus and right temporal pole, respectively, cause the largest decreases in the observed correlation (see Fig. 6). In other words, in subjects that learn quickly, the left hippocampus and right temporal pole seem to contribute to a higher separability dimension and \emph{vice versa}. A possible explanation for these results is that learning to perform this task requires effective separability of stimulus dimensions mediated by these regions. Such an interpretation is in line with the known role of the hippocampus in the rapid learning of stimulus associations \cite{squire1992memory}, and the role of the temporal pole in representing information about abstract conceptual properties of objects (such as object value) \cite{peelen2012conceptual}. In contrast, the removal of regions such as the left rostral middle frontal cortex and left supramarginal gyrus most strongly enhance the observed correlation, suggesting that their activity is orthogonal to or does not directly contribute to the large separability dimension that characterizes quick learners.

\subsection{The emerging relationship between dimension of neural data and response accuracy} 

\begin{figure}[h]
	\includegraphics[width=0.95\linewidth]{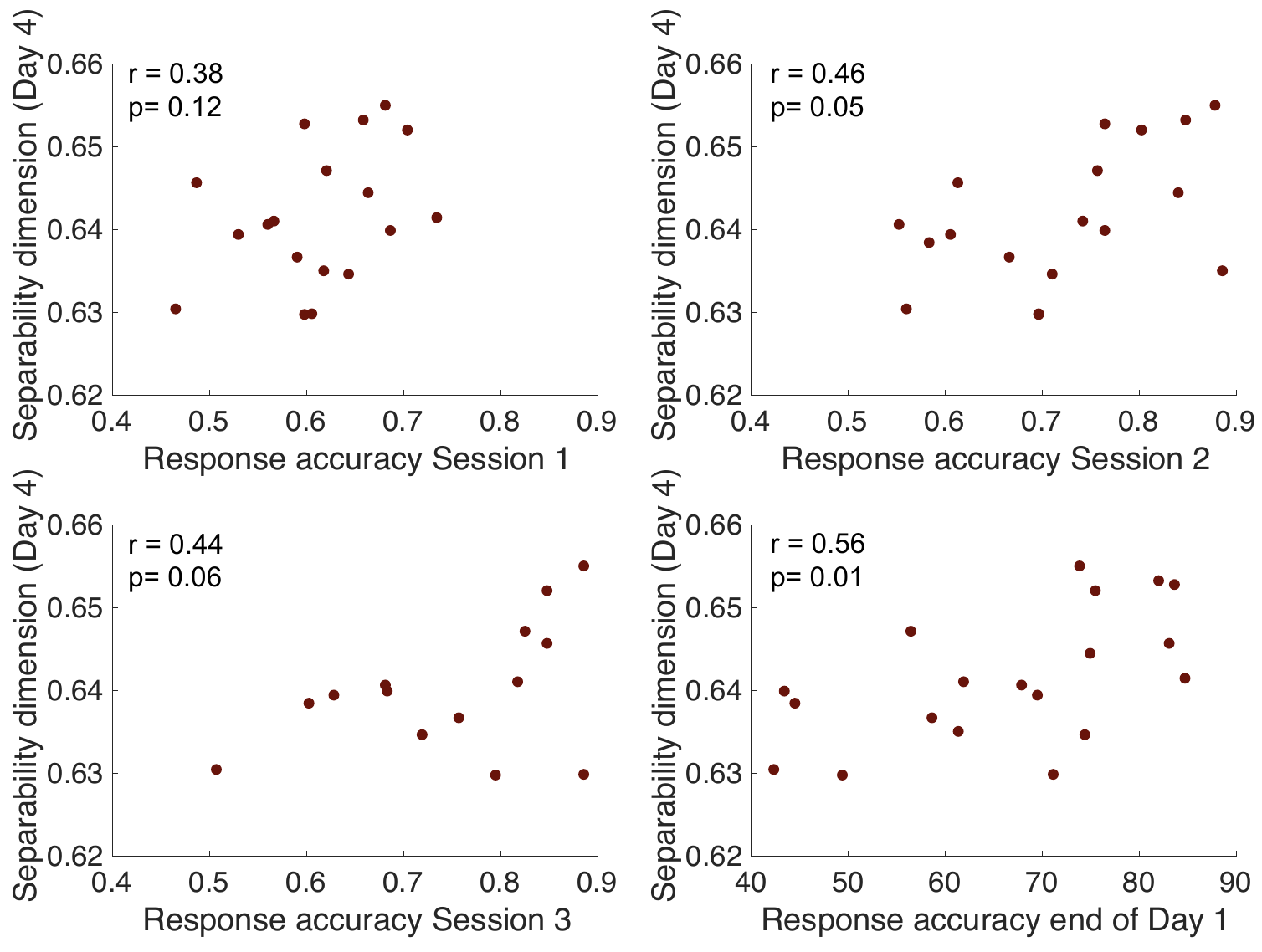}
	\caption{\textbf{Emerging relationship between dimension of neural responses and behavioral accuracy.}  We examine how performance accuracy changes across the three learning sessions and value judgement session on the first day of training where the greatest individual differences were observed, and its correlation with task-based separability dimension on the final day of training. We see that this correlation increases from $r=0.38$ in the first learning session (top left) to $r=0.56$ by the end of the first day in the value judgement session (bottom right). Note that in contrast to the non-parametric permutation test used to yield $p<0.001$ for the bottom right data in the main text, here we simply provide the parametric $p$-values from the Pearson's correlation which are much less computationally intensive to estimate}.\label{fig:day1}
\end{figure}

We also investigate how the learning of value emerges throughout the first day. We examine how the learning responses change across the three learning sessions and value judgement session on the first day, and ask whether individual differences in learning performance are correlated with their task-based separability dimension on the last day of training (see Fig. \ref{fig:day1}). We find that the correlation between performance and separability increases from $r=0.38$ in the first training session (Fig. \ref{fig:day1}, top left) to $r=0.56$ by the end of the first day in the value judgement session (Fig. \ref{fig:day1}, bottom right), suggesting that this relationship between the dimension of neural data and the response accuracy of participants emerges across sessions on the first day of training.

\subsection{Comparison with size judgment}

\begin{figure}[h]
\includegraphics[width=0.53\linewidth]{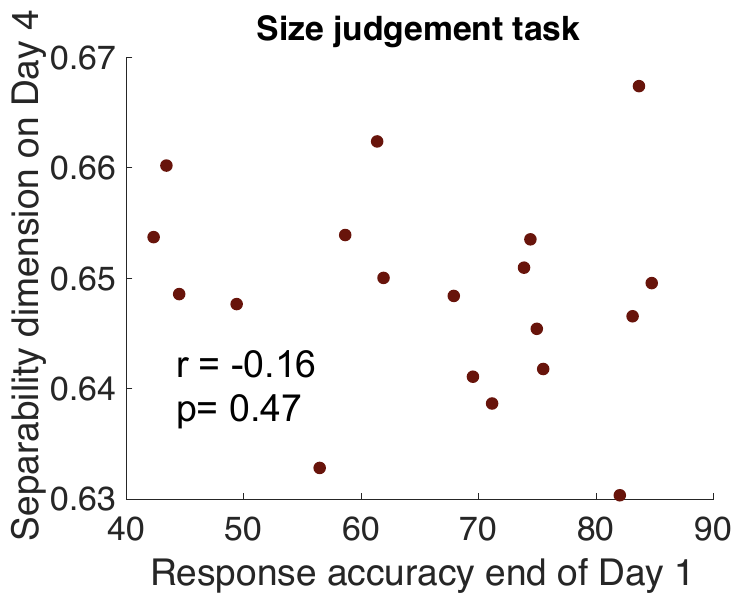}
\caption{\textbf{Dimension of neural data from size judgment session.} The separability dimension of data from the size judgment task on the last day does not show significant differences between quick and slow learners, suggesting that the cognitive task or effort of judging value itself is necessary for this emergence of a larger dimensional neural response.}\label{fig:size}
\end{figure}

We use our method on data from the size judgment session, which is very similar to the value judgment session in its setup and response format, but in which subjects are asked to evaluate the relative size of each shape instead of the relative value (see Methods). Unlike for the neural responses in the value judgment session from the same day, we find that quick learners do not show any differences in their task-based separability dimension. Indeed, this separability dimension of subject's representation during the size judgment task shows no significant correlation with the response accuracy of subjects (Pearson's correlation $r=-0.16$, $p=0.47$; see Fig. \ref{fig:size}). These results show that the representation of neural responses to various shapes has a larger dimension for quick learners only when they are asked to evaluate the relative value of these shapes. Hence, this larger dimensionality is not evoked purely by visual apprehension of these shapes, suggesting that the cognitive task or value judgement itself is necessary for this emergence of a larger dimensional representation.

\subsection{Changes in task-based separability dimension across the four days}

We track the separability dimension of neural data from the value judgment sessions held at the end of each day, to calculate their correlation with the response accuracy of participants on the first day. We find that there is little correlation between the separability dimension of neural data and the performance on the first day, $r=-0.05$, $p=0.85$, as compared to the fourth day, $r=0.56$, $p=0.01$, suggesting that this larger dimension of neural responses for quick learners also takes time to emerge (Pearson's correlations and parametric $p$-values reported). 

\begin{figure}[h]
\includegraphics[width=0.99\linewidth]{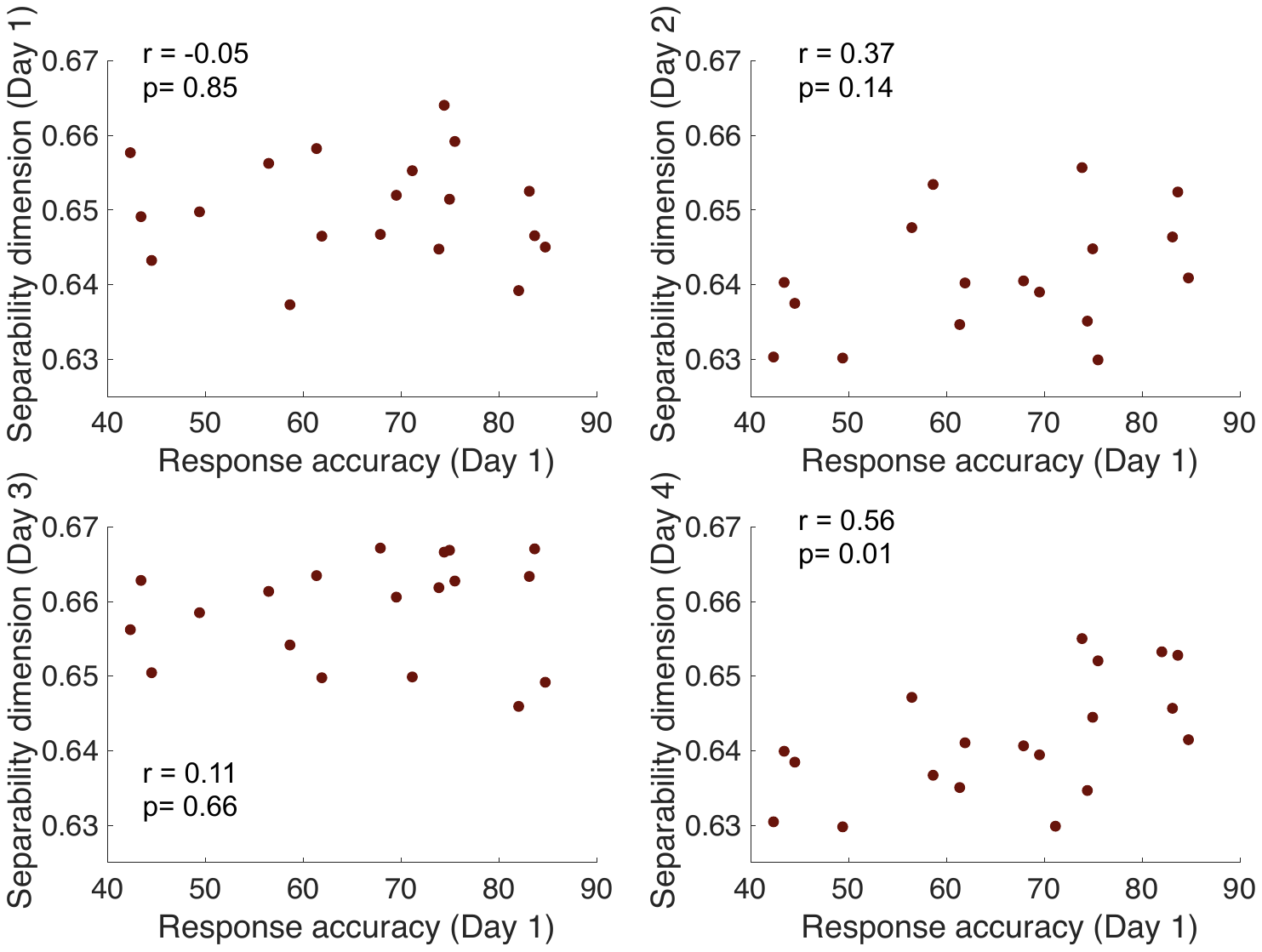}
\caption{\textbf{Changes in separability dimension across the four days.} We study the correlation of the separability dimension of neural data from the value judgment sessions at the end of each day, with the response accuracy of participants on the first day. We find that quick learners do not have a particularly large dimension of neural response patterns on the first day, $r=-0.05$, $p=0.85$, as compared to the fourth day, $r=0.56$, $p=0.01$, suggesting that this larger dimension for quick learners takes time to emerge. Pearson's correlations and parametric $p$-values reported here.}\label{fig:dimchanges}
\end{figure}

\begin{table}[h]
 \centering
\caption {\textbf{Results for all brain regions studied at the voxel level.} The left anterior cingulate passes non-parametric $p<0.005$ corrected for multiple comparisons (marked with $^*$).} 
\begin{tabular}{|c|c|c|c|c|}
\hline
No. of voxels & Brain region & Hemisphere & $r$ & $p$  \\
\hline
300 & Anterior cingulate & Left &  0.543 & 0.003$ ^*$ \\
300 & Anterior cingulate & Right &  0.306  &   0.063 \\
300 & Primary visual  & Left &  0.500  &   0.016 \\
300 & Primary visual  & Right &  0.090  &   0.390 \\
300 & Posterior fusiform & Left & 0.085& 0.396  \\
300 & Posterior fusiform & Right & 0.608 & 0.050 \\
300 & Lateral occipital & Left & 0.415& 0.092  \\
300 & Lateral occipital & Right & 0.0591&  0.465  \\
140 & Orbitofrontal cortex & Left & 0.142& 0.291  \\
140 & Orbitofrontal cortex & Right & 0.357& 0.103  \\
\hline
\end{tabular} \label{tab:voxelsall}
\end{table}

\newpage
\section{Replication of results}
\subsection{Different brain parcellation yields similar result}
\begin{figure}[h]
\includegraphics[width=0.99\linewidth]{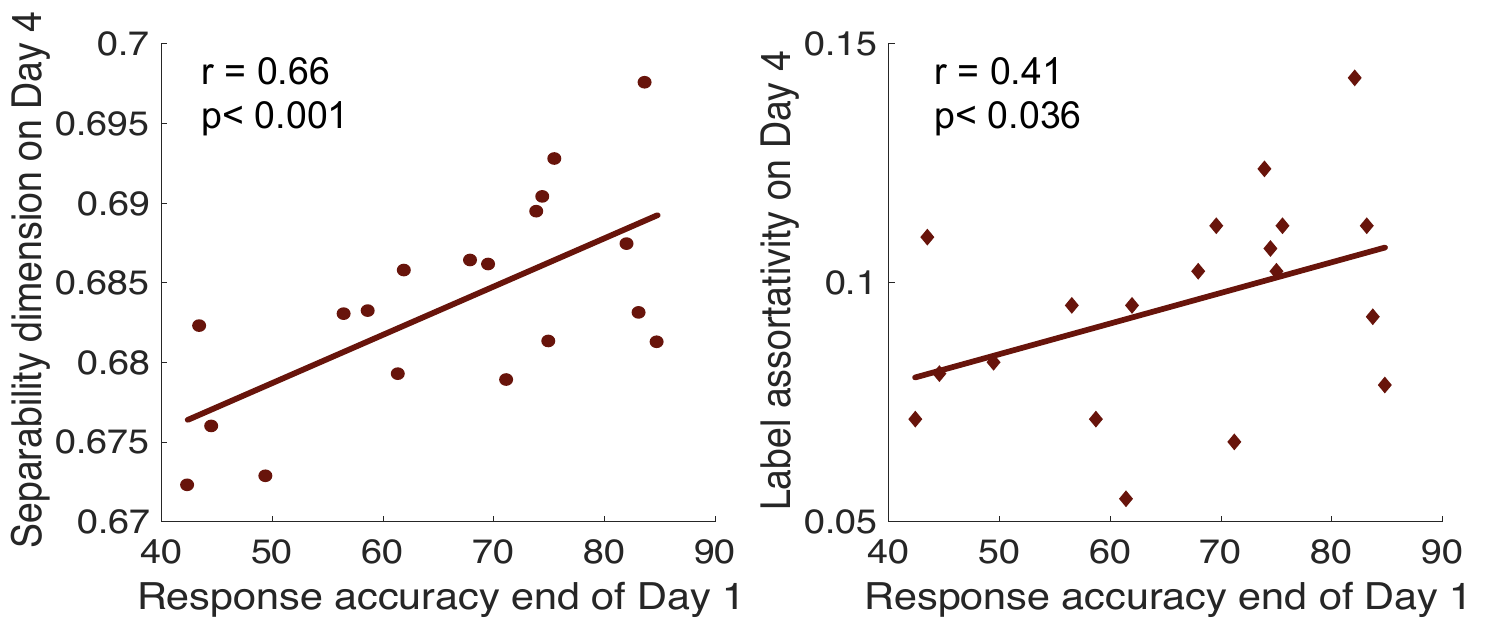}
\caption{\textbf{Replication of results using the whole-brain Power parcellation.} \textit{Left}: task-based separability dimension of a subject's representation on the fourth day is strongly correlated with the behavioral accuracy of subjects from the first day, with $r=0.66$ and non-parametric $p<0.001$ obtained from comparison with the null model. \textit{Right}: Label assortativity (retaining all twelve original labels) of the same data displays a positive trend with the response accuracy of subjects from the first day, with $r=0.41$ and non-parametric $p<0.036$ obtained from comparison with the null model. These results are consistent with our results obtained using the 83-region Lausanne parcellation.} \label{fig:replication} 
\end{figure}

We repeat our analyses on data obtained from a different whole-brain parcellation -- a functional-based parcellation that subdivides the brain into 264 regions \cite{Power2011}. Note that because not all subjects had data in 3 out of the 264 regions, we retain only the 261 brain regions with data for all participants. Upon repeating our calculations, we obtain similar results (see Fig. \ref{fig:replication}). 

\subsection{Quick learners have an increasing dimension of representation across the experiment}
\begin{figure}[h]
\includegraphics[width=0.53\linewidth]{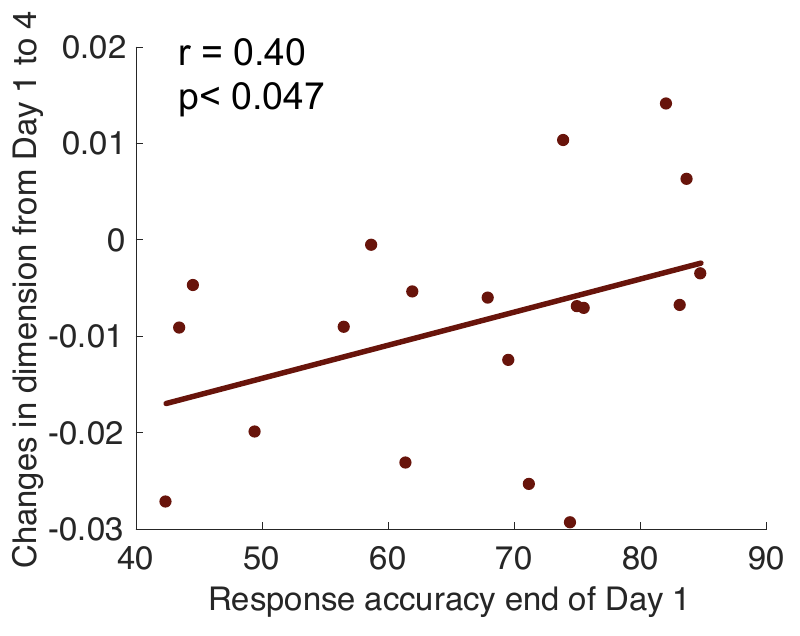}
\caption{\textbf{Quick learners have an increasing dimension of representation across the experiment.} We examine the changes in dimension of the task-based representation of each individual's neural data, from the first day to the last day. These changes are positively correlated with the subject's learning accuracy, consistent with our findings in the main analysis. A non-parametric permutation test shows that this correlation is significant with $p<0.047$.}\label{fig:neuralchanges}
\end{figure}

As our interests generally lie in understanding the process of learning, we are most interested in considering changes that occur during the full time course of the experiment. These changes are neatly and parsimoniously reflected by the outcomes of the learning process in terms of the neural representations on the final day. Thus, we focus the majority of our analyses on the neuroimaging data collected on this fourth and final day of training. An alternative approach is to consider changes in the neural data from the first day to the fourth day. Taking the changes in dimension of the geometric representation of each individual's neural data, we find that these changes are positively correlated with their learning accuracy ($r=0.40$; see Fig. \ref{fig:neuralchanges}). To verify that this correlation is statistically significant, we permute the differences among individuals to recalculate this correlation in 1000 bootstrapped samples, which yields $p<0.047$, confirming our findings from the main analysis. The consistency between the results of the two analyses is likely due to the fact that the neural data on the first day does not display a significant amount of variability between fast and slow learners (see Fig. 9 in this Supplemental document).

\subsection{Use of graded learning curve for behavioral metric confirms findings}
\begin{figure}[h]
	\includegraphics[width=0.53\linewidth]{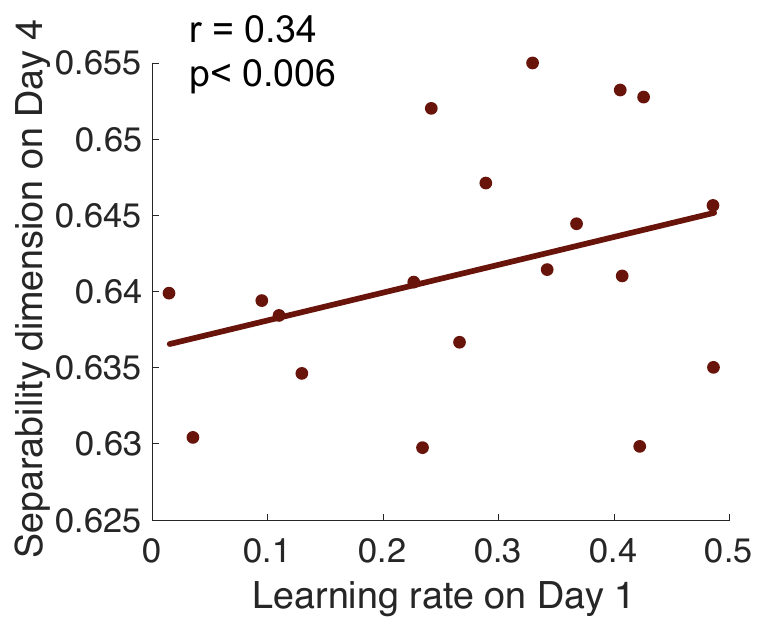}
	\caption{\textbf{Use of graded learning curve for behavioral metric confirms findings.} As an alternative measure of the learning rate for each subject, we use the slope of response accuracy across all three sessions of the learning phase in Day 1. We observe that this learning rate has a positive correlation with the dimension of representation across subjects, with a non-parametric permutation test yielding $p<0.006$, consistent with our findings in the main text. }\label{fig:learningrate}
\end{figure}
To verify that our results do not depend unduly on the specific choice of behavioral metric, here we use an alternative measure of learning speed. Specifically, we consider the slope of response accuracy across all three sessions of the learning phase on Day 1; this metric provides an estimate of the rate at which individuals learned to associate the assigned values to the presented shapes. To unpack this metric a bit further, we note that as each session consisted of 132 trials, where responses to each trial were binary (right or wrong), we examine the number of correct responses within a given window to give an average accuracy for that window. Windows of 22 trials were chosen in order to create 6 equally sized windows for each session. Hence, the three learning sessions on Day 1 yield 18 windows, and we calculate the slope of response accuracy across those windows for each individual. Next, we calculate the correlation between this slope (or learning rate) and the dimension of the task-based representation from day 4. We found that the two variables were positively correlated with one another ($r=0.34$; see Fig. \ref{fig:learningrate}). To assess the statistical significance of this correlation, we use a null model where the labels of each shape were permuted (the same null model as in the main text), and obtain a non-parametric $p<0.006$, confirming the findings that we report in the main text.

\end{document}